\documentclass[traditabstract]{aa}
\usepackage{txfonts}
\usepackage{natbib}
\usepackage[dvips]{graphicx}

\def\ltsima{$\; \buildrel < \over \sim \;$}
\def\lsim{\lower.5ex\hbox{\ltsima}}
\def\gtsima{$\; \buildrel > \over \sim \;$}
\def\gsim{\lower.5ex\hbox{\gtsima}}
\def\msun{~M_{\odot}}

\def\zsun{~Z_{\odot}}
\def\mdot {\dot M}

\begin{document}

\title
{Search for X-ray emission from subdwarf B stars with compact
companion candidates}
\author{S. Mereghetti\inst{1}, S. Campana\inst{2},  P. Esposito\inst{3},  N. La Palombara\inst{1},
A. Tiengo\inst{1,4} }

\institute{INAF, Istituto di Astrofisica Spaziale e Fisica
Cosmica Milano, via E.\ Bassini 15, I-20133 Milano, Italy
\and
INAF, Osservatorio Astronomico di Brera, via E.\ Bianchi 46,
I-23807 Merate (LC), Italy
\and
INAF, Osservatorio Astronomico di Cagliari, loc. Poggio dei
Pini, strada 54, I-09012 Capoterra (CA), Italy
\and
IUSS, Istituto Universitario di Studi Superiori, viale Lungo
Ticino Sforza 56, I-27100 Pavia, Italy}

\offprints{S. Mereghetti, sandro@iasf-milano.inaf.it}

\date{Received September 7, 2011 / Accepted October 24, 2011}

\authorrunning{S. Mereghetti et al.}

\titlerunning{}

\abstract {Stellar evolutionary models predict that most of the
early type subdwarf stars in close binary systems have white
dwarf companions. More massive companions, such as neutron
stars or black holes,  are also expected in some cases. The
presence of compact stars in these systems can be revealed by
the detection of X-rays   powered by accretion of the
subdwarf's stellar wind or by surface thermal emission. Using
the \emph{Swift} satellite, we carried out a systematic search
for X-ray emission from a sample of twelve subdwarf B stars
which, based on optical studies, have been suggested to have
degenerate companions. None of our targets was detected, but
the derived upper limits provide one of the few observational
constraints on the stellar winds of early type subdwarfs.   If
the presence of neutron star companions is confirmed, our
results constrain the  mass loss rates of some of these
subdwarf B stars to values $\mdot_\mathrm{W}<10^{-13}-10^{-12}
\msun$ yr$^{-1}$.

}

\maketitle

\section{Introduction}

Subdwarf B  (sdB) stars  are believed to be low mass stars
($\sim0.5 \msun$) with a helium-burning core surrounded by a
thin  hydrogen layer (see  \citealt{heb09} for a review of hot
subdwarfs). A large fraction of the known sdB stars are
in close binary systems \citep{max01,mor03}, supporting the
idea that   non-conservative mass transfer played a role in the
loss of the massive hydrogen envelopes of these stars.
\citet{han02} discussed different evolutionary channels leading
to the formation of sdB binaries. These computations predict
that the companions of sdB stars in short period systems
($\sim0.1-10$ days) should  fall into two main groups: late
type main sequence stars or white dwarfs. More rarely, if the
members of the original binary are massive enough, systems
composed of a sdB and a neutron star or black hole can also be
formed.

These predictions on the nature of sdB binaries are difficult
to test directly with optical observations. In fact most sdB
are only {\it single-lined} spectroscopic binaries, because
their   bright optical/UV emission outshines that from the much
fainter companions. On the other hand, the discovery and study
of hot subdwarfs   with compact companions is quite important,
since these systems might be among the progenitors of type Ia
supernovae \citep{ibe94}. Furthermore, the determination of the
mass and evolutionary stage of the two components, as done e.g.
for the binaries HD 49798 \citep{mer09,mer11} and KPD 1930+2752
\citep{gei07}, can shed light on  the poorly known processes
that take place during the common-envelope phases.

X-ray observations can help to identify systems
containing compact objects. X-rays can originate from surface
thermal emission of neutron stars or sufficiently hot white
dwarfs, or can be produced if the compact object accretes mass
from the sdB companion.
 Therefore, we carried out X-ray observations of a sample of sdB stars in
close binary systems, and for which evidence for a white dwarf,
neutron star, or black hole companion has been found.

\begin{table*}[htbp]
\caption{Main properties of the observed subdwarf binaries and
  results of \textit{Swift/XRT} observations.}
\begin{center}
\begin{tabular}{lcccccccc}
\hline
 Name          &  Period    &  Distance  &   $N_\mathrm{H}$ &  $M_\mathrm{C}$     &  Companion$^a$ &  Net Exposure Time & Count Rate$^b$ & $L_\mathrm{X}^c$  \\
               &(days)&(kpc)&($10^{20}$cm$^{-2}$)& ($\msun$)&      &   (s)       & (ct s$^{-1}$) & (erg s$^{-1}$) \\
\hline
\hline
\object{PG 1043+760}    & 0.12 & 0.66  & 3.5      & $>$0.1$^d$ & WD       & 5401  & $<$3.8$\times$10$^{-3}$ & $<$7.3$\times$10$^{30}$ \\
\object{PG 1432+159}    & 0.22 & 0.8   & 1.4      & 1.49--4.6  & NS/BH    & 3390  & $<$3.8$\times$10$^{-3}$ & $<$9.8$\times$10$^{30}$ \\
\object{PG 2345+318}    & 0.24 & 0.9   & 4.4      &  ...$^e$    & WD       & 3804  & $<$4.0$\times$10$^{-3}$ & $<$1.5$\times$10$^{31}$ \\
\object{HE 0532-4503} & 0.27 & 2.8   & 3.3      & 2.08--3.94 & NS/BH    & 4729  & $<$2.3$\times$10$^{-3}$ & $<$7.8$\times$10$^{31}$ \\
\object{CPD -64 481}  & 0.28 & 0.21  & 4.3      & 0.38--1.04 & WD       & 4572  & $<$3.8$\times$10$^{-3}$ & $<$7.9$\times$10$^{29}$  \\
\object{PG 1101+249}    & 0.35 & 0.39  & 1.0      & 1.09--2.44 & WD/NS/BH & 5263  & $<$2.8$\times$10$^{-3}$ & $<$1.7$\times$10$^{30}$ \\
\object{PG 1232-136}  & 0.36 & 0.57  & 3.4      & $>$6       & BH       & 5589  & $<$1.8$\times$10$^{-3}$ & $<$2.6$\times$10$^{30}$ \\
\object{GD 687}         & 0.38 & 1.1   & 2.0      & 0.50--0.93 & WD       & 5149  & $<$2.1$\times$10$^{-3}$ & $<$1.1$\times$10$^{31}$ \\
\object{HE 0929-0424} & 0.44 & 1.9   & 3.0      & 1.18--2.7  & WD/NS/BH & 5566  & $<$2.0$\times$10$^{-3}$ & $<$3.2$\times$10$^{31}$  \\
\object{PG 1743+477}    & 0.52 & 1     & 2.5      & $>$1.66    & NS/BH    & 3342  & $<$4.3$\times$10$^{-3}$ & $<$1.8$\times$10$^{31}$ \\
\object{PG 0101+039}    & 0.57 & 0.33  & 2.5      & 0.52--0.92 & WD       & 4851  & $<$2.3$\times$10$^{-3}$ & $<$1.0$\times$10$^{30}$  \\
\object{TON S 183}       & 0.83 & 0.54  & 1.2      & 0.63--1.33 & WD       & 4980  & $<$2.1$\times$10$^{-3}$ & $<$2.5$\times$10$^{30}$ \\
\hline
\end{tabular}
\end{center}

$^a$ WD = white dwarf, NS = neutron star, BH = black hole.

$^b$ 3$\sigma$ upper limit in the 0.3-10 keV energy range.

$^c$ In the 0.3-10 keV range, corrected for the absorption.

$^d$ The presence of a white dwarf in this system with high
inclination and short period is suggested by the non-detection
of a variable reflection component in the light curve
\citep{max04}.

$^e$ This sdB does not rotate synchronously, but the eclipse
observed in this system  indicates the presence of a white
dwarf companion \citep{gre04}.
\end{table*}

   \begin{figure*}
   \centering
   \includegraphics[width=\textwidth]{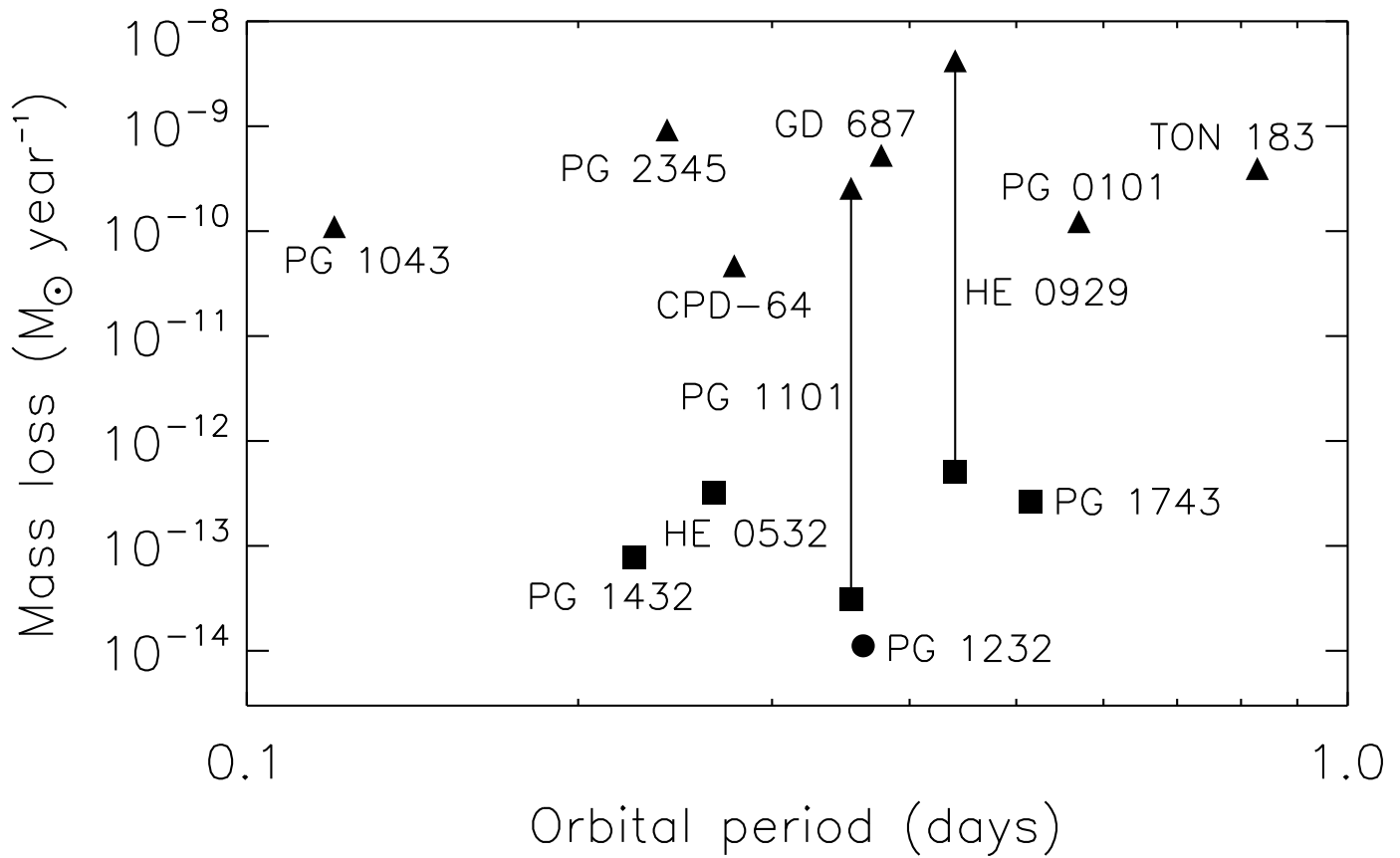}
      \caption{ Upper limits (3$\sigma$) on the sdB mass loss
rates in the case of a white dwarf (triangles), a
neutron star (squares), or a black hole (circle) companion.}
         \label{fig1}
   \end{figure*}
%

\section{Sample selection}

Several candidate  subdwarfs with compact companions have been
recently reported, based on radial and rotational velocity
measurements of single-lined spectroscopic binaries
\citep{gei10}. These authors showed that, with the reasonable
assumption that the sdB star rotates synchronously with the
orbital period, it is possible to derive the system
inclination, hence to set a lower limit on the mass of its
unseen companion.  In many cases the limits derived by
\citet{gei10} exceed the masses of late-type main-sequence
stars, implying the presence of a white dwarf or, in a few
cases, of a neutron star or a black hole.

Among the possible candidates, we considered those with the
lowest interstellar absorption ($N_\mathrm{H}<5\times10^{20}$
cm$^{-2}$), and with the shortest orbital period (0.1--0.8
days). The latter condition favors a higher X-ray luminosity
because in close binaries the compact object's orbit lies in a
region where the mass donor's wind is denser and slower, thus
giving rise to a larger accretion rate. Our targets are listed
in Table 1, where we give some of their properties. The values
for the estimated companion masses $M_\mathrm{C}$ (column 5)
are those derived from the analysis of \citet{gei10}. The
$N_\mathrm{H}$ values reported in Table 1 correspond to the
total Galactic column density\footnote{derived from the on-line
tool http://heasarc.nasa.gov/cgi-bin/Tools/w3nh/w3nh.pl, based
on the works of \citet{kal05} and \citet{dic90}.} in the source
directions,  and therefore they likely overestimate the actual
interstellar absorption of our targets.

\section{Observations and results}

The observations were carried out with the \emph{Swift/XRT}
telescope \citep{geh04} between May  and August 2011.   All the
data were obtained in photon counting mode. For most of our
targets the observations were split in a few short exposures,
carried out over a time period spanning several days. The net
exposure times, obtained by summing all the pointings of each
target, are  reported in Table 1.

No X-ray sources were detected in the \emph{Swift/XRT} images
at the positions of our targets. For each target we calculated
a 3$\sigma$ upper limit (following Gehrels 1986)
based on the exposure-corrected count rate measured at the
source position in the 0.3-10 keV energy range, taking also
into account the instrumental effects on the point-spread
function and the vignetting. The count rate upper limits were
then converted to X-ray fluxes, assuming a power law spectrum
with photon index 2 and the $N_\mathrm{H}$ values reported in
Table 1. The corresponding upper limits on the 0.3-10 keV
unabsorbed X-ray luminosities are typically in the range from
$\sim10^{30}$ to a few $\sim10^{31}$ erg s$^{-1}$. We used a
power law spectrum with the  assumption that X-rays are
produced by accretion. Thermal emission from neutron stars or
hot white dwarfs would yield a much softer spectrum, peaking in
the far UV/soft X-ray range,  with negligible contribution
above $\sim0.3$ keV where the \emph{Swift/XRT} is most
sensitive\footnote{For example, an extrapolation of the
spectrum measured below 0.25 keV for the hot white dwarf HZ43
\citep{beu06} would give a 0.3-0.5 keV count rate below 0.01 ct
s$^{-1}$ in the \emph{XRT}. Scaling  for the 3 to 40 times
larger distances of our targets and considering even a modest
interstellar absorption ($N_\mathrm{H}=10^{19}$ cm$^{-2}$,
compared to $\sim$10$^{18}$ cm$^{-2}$  for HZ43) yields count
rates between 10$^{-6}$ and $3\times10^{-4}$ ct s$^{-1}$.}. The
expected count rate in case of thermal emission, would strongly
depend on the interstellar absorption and on the star's surface
temperature and composition. Since the resulting upper limits
are not particularly constraining, also in view of the
distances of our targets, we restrict the following discussion
to the case of accretion powered emission.

\section{Discussion}

The lack of  X-ray detections of the sources in our sample does
not allow us to confirm the presence of compact stars in these
systems.  On the other hand,  in the hypothesis that a
collapsed object is indeed present in these binaries, our
negative result can be used to constrain the poorly known
properties of the sdB stellar winds. In fact, in all the
systems we considered, accretion onto the compact object must
occur through stellar wind capture, since  the sdB radius is
much smaller than the Roche lobe.

Little observational information is currently available on the
mass loss rate, $\mdot_\mathrm{W}$, from early type subdwarfs.
Evidence for $\mdot_\mathrm{W}\sim10^{-9}-10^{-8}$  $\msun$
yr$^{-1}$ has been inferred from UV spectroscopy of a few O
type subdwarfs \citep{ham81,ham10}. Lower values of
$\mdot_\mathrm{W}$ are expected in sdB stars \citep{ung08},
which are less luminous and have smaller temperatures than sdO
stars. \citet{vin02} derived an expression to compute the
expected mass loss rate for early type stars as a function of
their effective temperature, mass, luminosity, and metallicity.
With the appropriate parameters for each sdB star of our sample
(from Table 1 of Geier et al. 2010), and assuming solar
metallicity, this relation yields values of $\mdot_\mathrm{W}$
in the range $\sim10^{-12}-10^{-11}$ $\msun$ yr$^{-1}$.

Our luminosity upper limits can be converted in limits on
$\mdot_\mathrm{W}$ as follows. The expected mass accretion rate
onto the compact object, $\mdot$, can be simply estimated,
assuming Bondi-Hoyle accretion,  from the relation

\medskip
$\mdot_\mathrm{W} \pi R_\mathrm{a}^{2} = \mdot 4 \pi a^2$,
~~~~~~~~~~~~~~~~~~~~~~~~~~~~~~~~~~~~~(1)
\medskip

\noindent where $a$ is the orbital separation, and
$R_\mathrm{a} = 2GM/(v_\mathrm{o}^2 + v_\mathrm{W}^2)$  is the
accretion radius, which depends on the mass $M$ of the
accreting object, on its velocity $v_\mathrm{o}$, and on the
velocity of the stellar wind $v_\mathrm{W}$. We computed the
expected $\mdot$  for each source, using the parameters of
Table 1, a typical sdB mass of 0.45 $\msun$, and conservatively
assuming $v_\mathrm{W}$ equal to the sdB escape velocity. To
compute $R_\mathrm{a}$, and to convert luminosity to accretion
rate, we assumed for the white dwarf case  $M=0.6 \msun$ and
radius 10$^9$ cm, while for the neutron star case we used
$M=1.4 \msun$ and radius 10$^6$ cm. For PG 1232-136, in which
the optical mass function suggests the presence of a black
hole, we assumed a black hole of 6 $\msun$ with an accretion
efficiency of 10\%. The derived limits on $\mdot_\mathrm{W}$
are plotted in Fig. 1, where different symbols are used to
indicate the assumed compact object.

The limits we obtained for the systems likely hosting white
dwarfs (triangles in Fig. 1) are above
$\mdot_\mathrm{W}\sim5\times10^{-11} \msun$ yr$^{-1}$. Although
they are not particularly constraining for the sdB wind
properties, we note that they represent one of the few
observational results in this field. Deeper X-ray observations
of the closest candidates (e.g. PG 0101+039 and CPD -64 481)
with more sensitive instruments will be able to detect
accreting white dwarfs, if their sdB companions lose mass at a
rate $\mdot_\mathrm{W}\sim10^{-12}-10^{-11} \msun$ yr$^{-1}$,
as predicted by theoretical wind models.

More interesting constraints can be inferred from the binaries
likely containing neutron stars or black holes (PG 1432+159, HE
0532-4503, PG 1232-136, and PG 1743+477). Of course, an obvious
explanation for their non-detection is that some of the
assumptions made to infer the presence of compact objects is
wrong. We will not consider further this possibility, but note
that, as extensively discussed in \citet{gei10}, it is rather
unlikely. We also note that our simple estimate of the mass
accretion rate in these systems leads to conservative upper
limits on $\mdot_\mathrm{W}$. In fact, most wind-accreting
neutron stars in high mass X-ray binaries show an X-ray
luminosity larger than that predicted by Eq. 1. Therefore, we
are quite confident that the lack of X-ray emission from these
systems implies that the sdB stars have rather weak winds, with
$\mdot_\mathrm{W}<3\times10^{-13} \msun$ yr$^{-1}$.  This is
significantly below the predictions of the relation by
\citet{vin02}, if a solar metallicity is assumed.  A
metallicity $Z=0.3 \zsun$, or lower, is required for these sdB
stars to be consistent with the derived upper limits.

\section{Conclusions}

Using the \emph{Swift/XRT} telescope, we have carried out the
first systematic search for accreting compact object in
binaries with B type subdwarfs.  None of the sdB binaries in
our sample, selected on the basis of optical studies suggesting
the presence of collapsed companions, was detected in the X-ray
band, with typical luminosity upper limits of
$\sim10^{30}-10^{31}$ erg s$^{-1}$. If the presence of neutron
stars is confirmed in some of these systems, their lack of
X-ray emission implies sdB mass loss rates smaller than
$\mdot_\mathrm{W}\sim10^{-13} \msun$ yr$^{-1}$, below the
predictions of theoretical models for radiatively driven
stellar winds.

\begin{acknowledgements}

We thank Ralf Napiwotzki and the referee  Martin Barstow for
their useful suggestions.  We acknowledge the use of public
data from the \emph{Swift} data archive. This work was
partially supported with contributions from the agreements
ASI-INAF I/009/10/0 and I/032/10/0. PE acknowledges financial
support from the Autonomous Region of Sardinia through a
research grant under the program PO Sardegna FSE 2007--2013,
L.R. 7/2007 ``Promoting scientific research and innovation
technology in Sardinia''.

\end{acknowledgements}

\end{document}